\documentstyle[multicol,epsfig,aps,prb]{revtex}

\ifpreprintsty \def\multb{ } \def\multe{ } \else \def\multb{ \begin{multicols}{2}} \def\multe{ \end{multicols}} \fi

\begin{document}
\draft
\date{\today}
\title{ The origin of high transport spin polarization in La$_{0.7}$Sr$_{0.3}
$MnO$_{3}$: direct evidence for minority spin states}
\author{ B. Nadgorny$^{1}$, I.I. Mazin$^{1}$, M. Osofsky$^{1}$, R.J. Soulen$%
^{1}$, Jr., P. Broussard$^{1}$, R.M. Stroud$^{1}$, D.J. Singh$^{1}$, V.G.
Harris$^{1}$, A. Arsenov$^{2}$, and Ya. Mukovskii$^{2}$.}
\address{$^1$Naval Research Laboratory, Washington, 
DC 20375\\
$^2$Moscow Institute of Steel and Alloys, 
Moscow, Russia}
\maketitle

\begin{abstract}
{ Using the point contact Andreev reflection technique, we have carried out a
systematic study of the spin polarization in the colossal magnetoresistive
manganite, La$_{0.7}$Sr$_{0.3}$MnO$_{3}$} { (LSMO). Surprisingly, we
observed a significant increase in the current spin polarization with the
residual resistivity. This counterintuitive trend can be understood as a
transition from ballistic to diffusive transport in the contact. Our results
strongly suggest that LSMO does have minority spin states at the Fermi
level. However, since its current spin polarization is much higher than that
of the density of states, this material can mimic the behavior of a true
half-metal in transport experiments. Based on our results we call this
material a}  {\it transport} { half-metal.}
\end{abstract}
\pacs{}
\multb

A half-metallic ferromagnet is a metal that has an energy gap at the Fermi
level, {\it E}$_F$ , in one of the two spin channels. Only the other
channel has states available for transport, and thus the electric current is
fully spin-polarized. Finding half-metallic or other highly spin-polarized
metals would bring about major advances in magnetoelectronics, since device
performance improves dramatically as the spin polarization of the metal
approaches 100\%.\cite{5} Although half-metallicity has been predicted in
quite a number of materials, the experimental situation is still
controversial, especially for the manganese perovskite, La$_{0.7}$Sr$_{0.3}$%
MnO$_{3}$ { (LSMO). Theoretical \cite{6} and experimental values\cite%
{7,8,9,10} of the spin polarization of this fascinating material with highly
unusual structural, magnetic and electronic properties, obtained by different
techniques vary from 35\% to 100\%. Not surprisingly, when Park} {\it et al}%
{. concluded from their spin-resolved photoemission spectroscopy measurement
that LSMO is completely spin-polarized \cite{7}} it attracted immediate
attention. This result was important not only from a practical viewpoint,
but also as a potential new insight into the microscopic physics of this
system, since the values of the spin polarization are extremely sensitive to
the band structure of LSMO.\cite{DE} The conclusion of Ref.\onlinecite{7}{, however,
disagrees with the band structure calculations \cite{6}} which predicted
only 36\% for the Fermi level density-of-states (DOS) spin polarization \cite%
{11} of the bulk La$_{0.7}$Ca$_{0.3}$MnO$_{3}$(La$_{0.7}$Sr$_{0.3}$MnO$_{3}$%
). Spin-resolved tunneling \cite{12}  experiments also indicate {\cite{8,9}%
} incomplete (54\% and 81\%, respectively) spin polarization for La$_{0.66}$%
Sr$_{0.34}$MnO$_{3}${. Recent LSMO-superconductor tunneling experiments
produced a spin polarization of 72\% \cite{10}. To address this controversy,
we have done systematic measurements of the transport spin polarization in La%
$_{0.7}$Sr$_{0.3}$MnO$_{3}$}  single crystals and thin films using the Point
Contact Andreev Reflection (PCAR) technique. 

Importantly, the measured value of the spin polarization, $P_{n}$,
depends on the experimental technique. It is often possible \cite{13} to
define {$P_{n}$ {in the following form: 
\begin{equation}
P_n=\frac { \langle N_\uparrow (E_F)v^n_{F\uparrow}\rangle- \langle
N_\downarrow (E_F)v^n_{F\downarrow}\rangle} { \langle N_\uparrow
(E_F)v^n_{F\uparrow}\rangle+ \langle N_\downarrow
(E_F)v^n_{F\downarrow}\rangle} 
\end{equation}
{where} $N_\uparrow ( E _{F}$), $N_\downarrow ( E _{F}$) and} $%
v_{F\uparrow}$, $v_{F\downarrow}$ {are the majority and minority spin DOS
and the Fermi velocities, respectively. This definition allows a direct
comparison between different experiments and the theory, since all the
quantities in Eq. 1 can be evaluated from the band structure. The spin
polarization} {{\it P}$_{0}$} {({\it n}} =0) measured by spin-resolved
photoemission measurements is determined only by the DOS at the Fermi level.%
\cite{14} Transport experiments measure a different spin polarization, which
includes the Fermi velocities (Eq.1). In the ballistic, or Sharvin, limit (mean
free path, { {\it L}, larger than the contact size,} { {\it d}) the DOS
is weighted linearly with} { {\it v}$_{F}$, and} { {\it P}$_{1}$} { is
measured.\cite{15} In the diffusive, or Maxwell regime ($L< d$), as in the
classical Bloch-Bolzmann theory of transport in metals, the weighting is
quadratic in} { {\it v}$_{F}$} { ({\it n}=2) and} { {\it P}$_{2}$} { is
measured (assuming that the transport relaxation time,} ${\tau}${ , is a
constant). Tunneling experiments probe yet another spin
polarization,} { {\it P}$%
_{T}$, which may still be formally defined using Eq.1 (for} { {\it n}} { =
2) by replacing the velocities with spin-dependent tunneling matrix elements.%
} It can be shown \cite{13} that for the simplest case of a specular tunnel
barrier with low transparency {\it P}$_{T}$ reduces to {\it P}$_{2}$. }

{ One can immediately see from Eq.1 that} { {\it P}$_{0}$,} { {\it P}$_{1}
$,} { {\it P}$_{2}$, and} { {\it P}$_{T}$} { can be dramatically
different. In LSMO, for instance, band structure calculations predicted} { 
{\it P}$_{0}$=36\%, whereas}  {\it P}$_{2}$=92\%. Since the bulk current
is proportional to $\langle N(E_F)v_F^2\rangle$,  { {\it P}$_{2}$=92\%
implies that
spin-majority electrons carry 96\% of electric current. A system where the
current is (nearly) fully spin polarized can be called a} { {\it transport}}
{ half metal, as opposed to a conventional half metal where} $N_\downarrow
( E_{F})=0$ and thus $P_{0}= P _{1} = P_{2} = P _{T}$ =100\%. Our paper
reports the first experimental observation of this effect.

{ Recently, Soulen} { {\it et al}} \cite{16} and Upadhyay { {\it et al}} %
\cite{17} introduced the use of Andreev reflection for measuring transport
spin polarization. The Andreev process\cite{18} allows propagation of a
single electron with energy below the superconducting gap $\Delta$ { from a
normal metal to a superconductor by reflecting at the interface as a hole
with the opposite spin.} In a non-magnetic metal this process is always
allowed, because every energy state of a normal metal has both spin-up and
spin-down electrons. { In a ferromagnet, however, the spin-up and spin-down
symmetry is broken, and Andreev reflection is limited by the number of
minority spin conductance channels.\cite{19} The measured degree of
suppression of Andreev reflection can be then directly related to the spin
polarization of the ferromagnet using an appropriately modified {\cite{16,20}%
} standard theory.\cite{21} This procedure
allows a quantitative determination of the
transport spin polarization of ferromagnetic materials.

Using this approach, we have studied thin films and bulk crystals of LSMO.
The films were grown on (100)-oriented NdGaO$_{3}$, MgO, and LaAlO$_{3}$} {\
substrates by off-axis sputtering \cite{22}}  and by pulsed laser
deposition. The growth conditions (substrate temperature and deposition
rate) were also varied to fabricate films of the same composition but with
different defect concentrations, and thus different residual resistivity.
The composition was determined by X-ray fluorescence with an accuracy of
5\%. The crystals were grown by a floating-zone technique.\cite{23}

The adjustment mechanism used for the PCAR measurements and the experimental
setup equipped with the standard electronics for tunneling measurements in
the temperature range between 1.5 K and 4.2 K are described elsewhere \cite%
{16}{. Sn tips were used for all the measurements reported here. Generally,
at least ten point-contact junctions were made for each sample, where the
contact resistance} { {\it R}$_{n}$} { was kept within the limits 100 }$%
\Omega >R_{n}>1\ \Omega $, as prescribed by Ref.\onlinecite{24}{. Normalized
conductance,} { {\it G}({\it V})/{\it G}$_{n}$, was then calculated using} {%
 {\it G}$_{n}$} { obtained for voltages} $V\ll \Delta /e${ . Each
normalized curve was then fitted with the model (see caption to Fig.1) to
obtain the magnitude of the spin polarization. }

As a further test of our technique, we measured the conductance { {\it G}(%
{\it V})/{\it G}$_{n}$} { for a single contact as we cooled the LSMO
samples through the superconducting transition temperature of Sn,} { {\it T}%
$_{c}$=3.7 K. Naturally, a strong temperature dependence of the conductance
was observed as the gap opened up (Fig. 1).} Each of {\it G}({\it V})/{\it G}%
$_{n}$ curves was then fitted independently using a modified BTK model with
only two adjustable parameters: spin polarization, {\it P} and the barrier
strength, {\it Z}. The value of the superconducting gap $\Delta (T)$ was
determined separately from the BCS dependence. Importantly{, the values of}
{ {\it P}} { for the same sample were practically independent of} { {\it T%
}, as expected for this temperature range}, $T\ll T{_{Curie}}$ { (see the
inset in Fig. 1). }

Having confirmed the consistency of our technique, we measured the spin
polarization in a number of La$_{0.7}$Sr$_{0.3}$MnO$_{3}$ { thin films and
bulk single crystals, whose residual resistivity ranged from 40} ${\mu }{%
\Omega }$ cm to 2000 ${\mu }{\Omega }${ cm (see Fig.2). Surprisingly, the
transport spin polarization was greater for samples with larger residual
resistivity. If the material were a true half metal, one might expect the
opposite trend: better samples would have less spin-flip scattering and thus
show spin polarization closer to 100\%. The observed result can be
understood, however, if we take into account the dependence of the transport
spin polarization on the ratio of the electronic mean free path to the
contact size, as discussed above. It is natural to expect that all values of
the transport spin polarization,} { {\it P}} { measured for our samples
should be confined between} { {\it P}$_{1}$} { (pure ballistic limit) and} 
{ {\it P}$_{2}$} { (pure diffusive limit). Using the values of the
densities of states $[N_{\uparrow }(E_{F})=0.58$ states/eV} Mn, $%
N_{\downarrow }(E_{F})$=0.27 states/eV Mn] and Fermi velocities ($%
v_{F\uparrow }=7.4\cdot 10^{5}$ { m/s,} $v_{F\uparrow }=2.2\cdot 10^{5}$ {\
m/s), from Ref.}\onlinecite{6},  we obtain { {\it P}$_{1}$} { = 74\%, and} { 
{\it P}$_{2}$} { = 92\% in fairly good agreement with the experimental
data. The last number implies that only 4\% of the current is carried by the
spin-minority channel, so we can assume for an estimate that the whole
conductance is due to the spin-majority channel. Using Ziman's expression
for conductivity,}
\begin{eqnarray}
\sigma_\uparrow & =&(1/3)e^{2}N_{\uparrow }(E_{F})v_{F\uparrow}^{2}\tau_\uparrow,
\end{eqnarray}
 { we obtain for the three values of resistivity,} ${\rho }\sim
40\ \mu \Omega $ cm, ${\rho }\sim 400\ \mu \Omega $ cm, and ${\rho }\sim 2000\ \mu
\Omega $ cm, the mean free paths, $L\sim 65$ \AA, $L\sim 6.5$ 
\AA, and
$L\sim 1.3$ \AA, respectively. The contact size can be estimated from the normal
resistance of the contact, { {\it R}$_{n}$. Using a general expression \cite%
{25} , we can express} { {\it R}$_{n}$} { in the following approximate
form:} 
\begin{equation}
R_{n}\approx \frac{4}{3\pi }\frac{\rho L}{d^{2}}+\frac{\rho }{2d}
\end{equation}
where $\rho $ is the residual resistivity. From Eq.3 we can find the contact
size {\it d} for given values of $\rho $ and {\it R}$_{n}$. For the lowest
residual resistivity samples with $\rho \sim 40\ \mu \Omega $ cm, we obtain $%
d\sim 35$ {\AA }. Therefore, these samples are in the ballistic regime and
the measured values of the transport spin polarization should correspond to 
{\it P}$_{1}$. The resistivity range ${\rho }\sim 400\ \mu \Omega $ cm
corresponds to the intermediate regime $(L\sim d)$, whereas the highest
resistivity samples ${\rho }\sim 2000\ \mu \Omega $ cm are in the diffusive
regime ($L\ll d$), consistent with our measurements.

The theory of Ref. \onlinecite{21}, as well as its modified version \cite{16}, is
directly applicable only to the ballistic transport case ($L\gg d$). The
complete theory for an arbitrary transport regime has yet to be developed.
We did derive expressions for purely diffusive regime, $L\ll d$, which will
be published elsewhere.\cite{20} Importantly, when our high residual
resistivity samples data were fitted with these expressions, we found that
the only appreciable change was in the values of {\it Z}, while the spin
polarization remained basically unchanged.

The analysis described above is based on the assumption that higher
resistivity of our samples is mainly due to the increase of the number of
defects and corresponds to shorter mean free path. However, higher
resistivity of our samples could have also resulted from the
presence of grain boundaries or some other extraneous effects. To make sure
that the residual resistivity in LSMO is, indeed, controlled by the
concentration of defects, we performed extended X-ray absorption fine
structure (EXAFS) measurements that directly probe the local structure of a
material. Specifically, the measurements on the Mn K-absorption edge give a
quantitative description of the real-space local environment around the Mn
cations, allowing us to reconstruct the MnO$_{6}$ octahedra in LSMO \cite%
{26,27}. Three films and the single crystal samples with the resistivity $%
40\ \mu \Omega $ cm $<\rho <800\ \mu \Omega $ cm were measured. We found that
for all these samples, the MnO$_{6}$ octahedra experience little or no
distortion. On the other hand, the measurements also indicate that the
Mn-La/Sr bond length changes from site to site, likely due to La/Sr site
defects, which are known to occur in LSMO. These differences are seen in the
Mn-La/Sr correlation, where the amplitude of the Fourier peak systematically
decreases with increasing residual resistivity. The change in the amplitude of
this peak (while the amplitude of the nearest neighbor O peak remains
unchanged) is consistent with an increase in A-site cation defects with
increasing residual resistivity. This result is in agreement with the neutron
diffraction refinement of LSMO samples processed under a variety of
conditions that indicate the propensity for A-site defect formation.\cite{28}
Such defects lead to tilting and/or rotation of the octahedra, without
necessarily introducing any local distortions. Defect concentrations
estimated from the EXAFS correlate with the residual resistivity,
demonstrating that the latter is due mainly to electron scattering by
defects.

Additionally we took one of our low resistivity films and irradiated it with
10 MeV Si ions, which increased the residual resistivity. This allowed us to
measure the change in the spin polarization as a function of residual
resistivity in the {\it same} sample. We have found that the spin
polarization for the irradiated sample follows the same curve (shown in Fig.
2) as for as-grown films and crystals. Although the defects in these two
cases may not be of the same nature, they apparently affect the scattering
rates in LSMO similarly, at least within a limited defect concentration
range. 

The EXAFS results analysis together with the irradiation experiment
allow us to conclude that the observed universal dependence of the spin
polarization on the resistivity is directly correlated with the carrier mean
free path in LSMO and is not determined by the surface or morphology of the
samples.

It is interesting to note that the conductivity is mostly ( $\sim $95\%)
determined by the spin-majority band, while the spin polarization is
controlled by the minority band. Moreover, because of the large disparity
between the two bands, the same defects are likely to influence the
transport in the minority band stronger, as it is much easier for a minority
band to approach the minimal metallic conductivity limit $k_{F} L\sim 1$. In
this case, defects will dramatically modify the minority carrier properties,
without significantly affecting the majority carrier properties. Ultimately,
the minority carriers can be localized by the disorder, with the majority
carriers retaining a long mean free path, to maintain overall metallic
conductivity, in which case the transport spin polarization will approach
100\%. Thus localization effects may be viewed as a limiting case of a
transition from the ballistic to the diffusive regime.
It is important to emphasize the difference between the 
spin polarization of the current through the interface and 
the spin polarization of the bulk current.
While the former can change from the ballistic limit, $P_1$, to the
diffusive limit, $P_2$, and possibly eventually to 100\% if the
minority carriers become fully localized, the latter is defined
by the diffusive formulas even for the cleanest samples. Of course,
localization affects the bulk transport spin polarization as well as the
contact spin polarization. 

The results of our transport spin polarization measurements for LSMO are
consistent with the tunneling measurements of Refs. \onlinecite{8,9,10} and the
band structure calculations \cite{6}, demonstrating that this material is 
{\it not} a half metal. The agreement with Ref. \onlinecite{6} is in fact quite
remarkable, considering the approximations used in the calculations (Local
Density Approximation, perfect La/Sr ordering). Some discrepancy
between the theoretical
prediction for the ballistic limit (74\%) and our experimental values ($\sim 
$60\%) is therefore
not surprising (in the diffusive limit the agreement is
almost perfect, $P\sim $90\%).

How can our results be reconciled with the
100\% polarization inferred from photoemission \cite{7}? First, we note that
our lowest resistivity films (and single crystals) are almost identical to
the sample described in Ref. \onlinecite{7}. Both have residual resistivity of $%
40\ \mu\Omega$ cm$<{\rho}< 800\ 
\mu\Omega$ cm, Curie temperature of $\approx 350$
K and coercive force of $\approx$ 50 G (see Fig.3). On the other hand, the
band structure calculations, which agree well with our measurements, predict
for the photoemission-probed spin polarization, {\it P}$_{0}$, a low number
of $ 36$\%. Since only $\sim $1 nm surface layer is accessible to the
photoemission \cite{7}, one possible resolution of this discrepancy would be
that only the surface of the sample, which had undergone a complex cleaning
procedure \cite{7}, was half-metallic.\cite{29}  Indeed, it is well known
that one of the main surface effects on the electronic structure is the
overall band narrowing, as a surface atom has a smaller coordination number, 
{\it Z}, than a bulk atom (by $\approx 20$\% for the cubic perovskite
lattice). Thus the overall bandwidth, which is roughly proportional to the
product {\it Zt} ({\it t} being the effective hopping), is reduced at the
surface by the same amount.\cite{31} As the minority band in La$_{0.7}$Sr$%
_{0.3}$MnO$_{3}$ is quite narrow and its edge is very close ($\approx$ 0.2
eV) to the Fermi energy, band narrowing can easily make the system
half-metallic. To check whether a band narrowing of the order of 20\% may be
responsible for the results of Ref. \onlinecite{6}, we considered another problem,
which also has band narrowing, albeit for a different reason. Namely, we
calculated (in virtual crystal approximation) the effect of the uniform
lattice
expansion on the band structure of LSMO. This effect reduces {\it t} without
changing the coordination number. We have found that just 3\% linear expansion, which
corresponds to approximately 10\% reduction \cite{30} in {\it t}, corresponding
to 
10\% reduction in the overall bandwidth, is already
sufficient to make the
system half-metallic\cite{surf}. Therefore, it is quite plausible that the surface
layer of LSMO is, indeed, half-metallic.

Another possible explanation of the photoemission results can be attributed
to the spin-filtering, whereby a faster scattering rate for minority spin
electrons compared to the majority spin electrons leads to an excessive
apparent spin polarization.\cite{32}  This explanation is also consistent
with the lack of dispersion {\it E}({\it k}) observed in Ref.
 \onlinecite{7}.

{ In conclusion,} { our results unambiguously show the presence of the
minority electrons at the Fermi level in the bulk of LSMO,} { in good
agreement with the band structure calculations\cite{6},
{ indicating that this material is not a half-metal.} { At
the same time our measurements have directly shown a high degree of the
current spin polarization (58\%$<P{<}92$\%) in bulk LSMO. This result
confirms that this material is a promising candidate for magnetoelectronics
applications, since ultimately it is the current spin polarization that
controls the performance of these devices. One can call La$_{0.7}$Sr$_{0.3}$%
MnO$_{3}$, therefore, a} { {\it transport}} { half-metallic ferromagnet,
due to the fact that the spin polarization of the current in this material
is approaching 100\% in the high resistivity limit. The origin of this high
current spin polarization is, however, entirely different from that in the
case of a conventional half-metal. It is mostly due to the large difference
in the mobility of the spin-up and spin-down electrons, rather than their
DOS. Our conclusions are based not just on the measured spin-polarization
values themselves, but rather on the observation of a clear
correlation between the bulk resistivity and the measured spin polarization
which is} { {\it opposite}} { to the one expected for a true half-metal.\cite{33}
This picture agrees quite well with the band structure calculations, and the
results of the tunneling studies for this complex and remarkably rich
material system. }

The authors are grateful to T. Geballe, D. Worledge, and  I. Zutic
 for useful  discussions.

}\multe
\begin{figure}[tbp]
\centerline{\epsfig{file=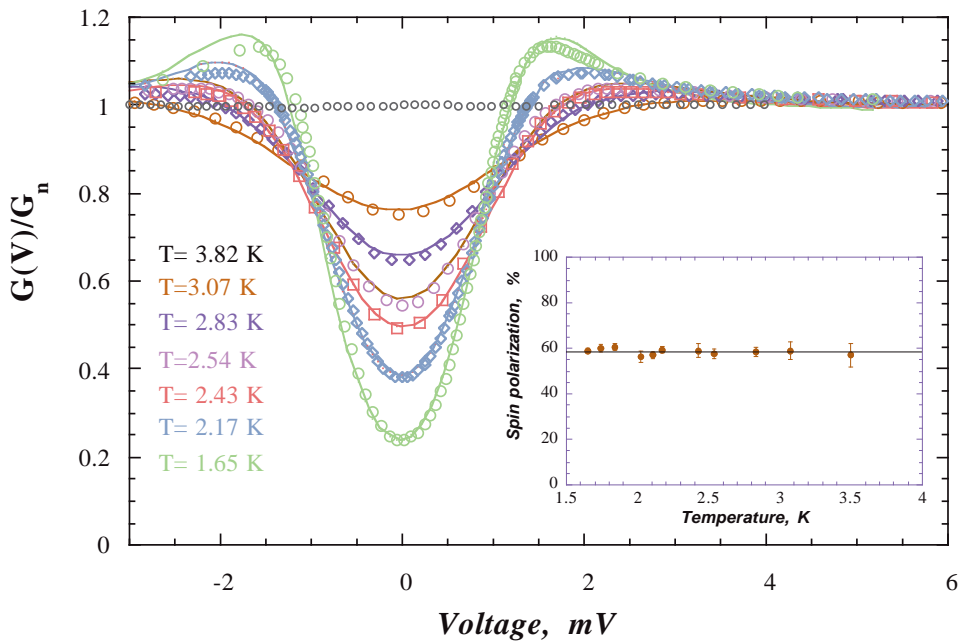,width=.8\linewidth,angle=90}}
\vspace{.1in} \setlength{\columnwidth}{6.5in} \nopagebreak
\caption{ Experimental data and fits at different temperatures for a La$%
_{0.7}$Sr$_{0.3}$MnO$_{3}$ film. The experimental data were corrected for
the lead resistance and small non-linearity in I-V characteristics above the
gap. The normalized  {\it G}({\it V})/{\it G}$_{n}$  curves were fitted
for all temperatures, varying the spin polarization,  {\it P}  and the
barrier strength,  {\it Z}. The BCS temperature dependence for the
superconducting gap $\Delta ( T )$ was used. Inset: Temperature dependence
of the spin polarization values for the same sample for 1.6 K${<} T{<}4.2 $
K. }
\end{figure}
\vspace{.2in}
\begin{figure}[tbp]
\centerline{\epsfig{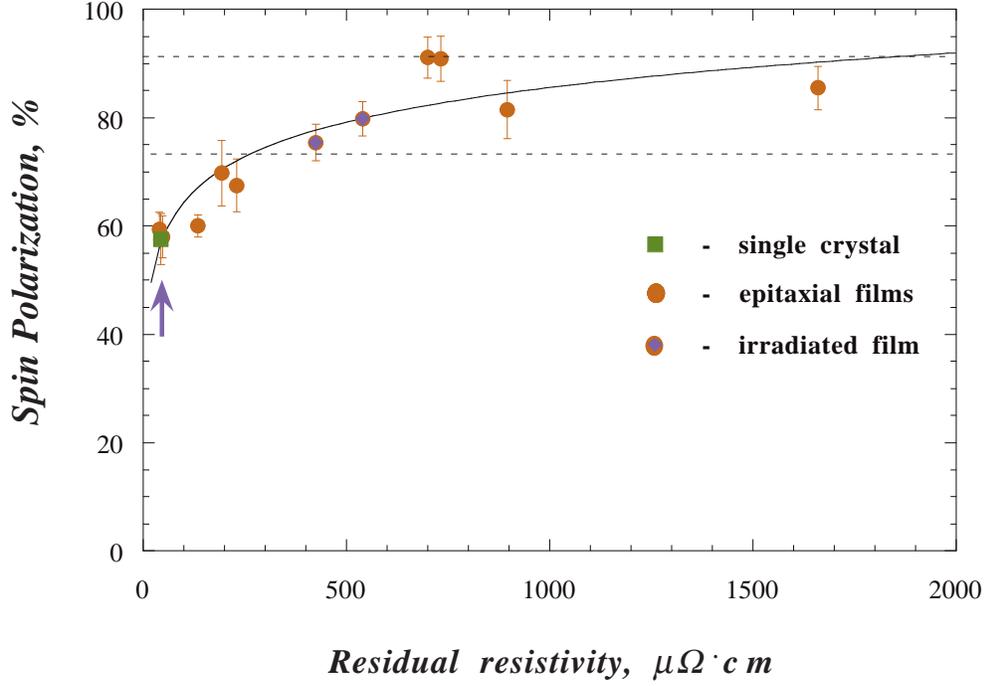}}
\vspace{.1in} \setlength{\columnwidth}{6.5in} \nopagebreak
\caption{ Spin polarization as a function of the residual resistivity of La$%
_{0.7}$Sr$_{0.3}$MnO$_{3}$ films and single crystals at $T$=1.6 K.
 The arrow indicates the lowest resistivity film that was later irradiated
 with Si ions (the higher resistivity film corresponds to the higher dose).
 Dashed lines correspond to $P_1$=74\% and $P_2$=92\%
 (see text). Solid line is a guide to the eye.}
\end{figure}
\begin{figure}[tbp]
\centerline{\epsfig{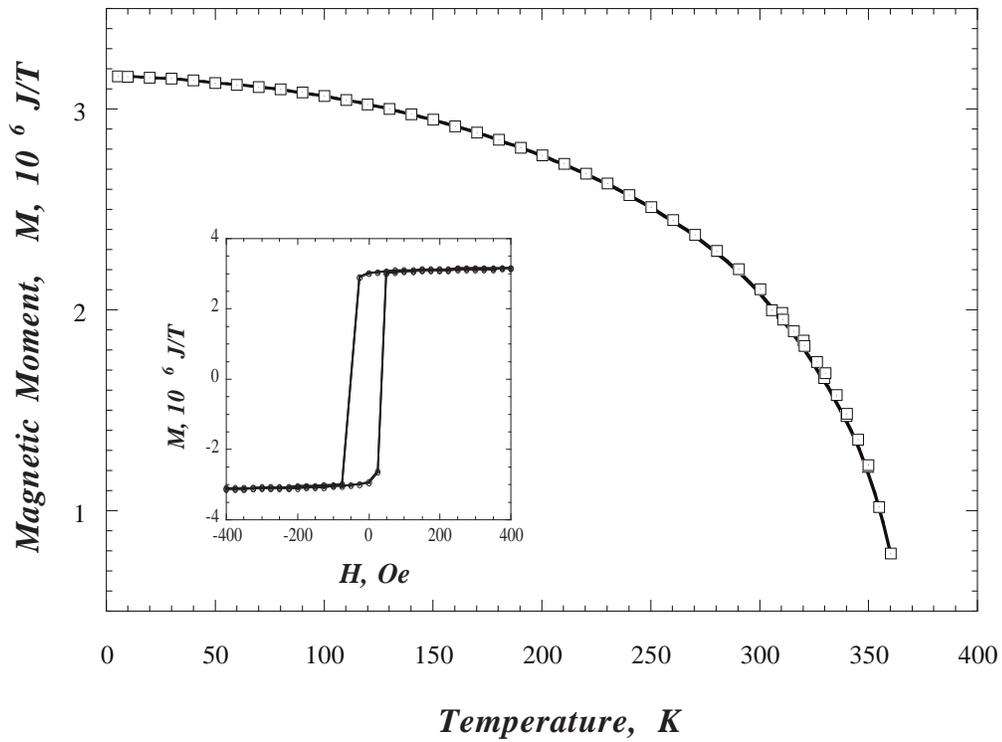}}
\vspace{.3in} \setlength{\columnwidth}{6.5in} \nopagebreak
\caption{ Temperature dependence of the magnetic moment of the lowest
resistivity film in the external in-plane field of 500 Oe. Inset: Hysteresis
loop for the same film at $T$=4.2 K.
}
\end{figure}


\begin{references}
\bibitem{DE} In an idealized double exchange system Hund's rule coupling
is much stronger than the hopping amplitudes, and hence the exchange
splitting is much larger than the bandwidth. Obviously, such a system
 is always a half-metal.
 LSMO, however, is far from this idealized picture,
and the two energy scales are comparable. In such a situation, whether
or not the material is half-metallic depends intimately on the details
of its electronic structure.
\bibitem{11} In Ref. \cite{6} only La$_{2/3}$Ca$_{1/3}$MnO$_{3}$ compound
was considered. We have done a few additional calculations for La$_{2/3}$Sr$%
_{1/3}$MnO$_{3}$ and found that substituting Sr for Ca practically does not
change the results.
\bibitem{26} The fine structure data extending from $\sim $30eV to 800 eV
beyond the Mn absorption edge { w}ere collected at room temperature in
fluorescence mode using a dynamic flow N gas ionization detector equipped
with Soller slits and a 6 $\mu $m Cr filter. Measurements were performed
using beam line X23B at the National Synchrotron Light Source (Brookhaven
National Laboratory, Upton, NY) with a storage ring current of 200-250 mA
and an energy of 2.5 GeV. EXAFS data were analyzed following standard
procedures leading to the Fourier transformation. Modeling of the near
neighbor Mn-O bonds was performed using the FEFF simulation codes of Ref. 
\onlinecite{27}.
\bibitem{29} As PCAR is probing the depth of the order of the mean free path
it is much less sensitive to the surface electronic structure.
\bibitem{33}  It is probably impossible to measure exactly 100\% spin
polarization using the PCAR technique even in a true half metal, due to 
potentially always-present spin-flip processes. However, the normalized
conductance, $G(0)/G(V)$, measured in our ``defect-free'' low resistivity
LSMO sample $G(V)/G(0)$(corresponding to $\sim $60\% spin polarization)
differs by a factor of 5 from $G(0)/G(V)$ for the sample with the highest
polarization measured to date by this technique ($\sim $95\% in CrO$_{2}$),
indicating a profound difference between the two materials.
\end{references}
\end{document}